# Dry dilution refrigerator with He-4 precool loop

Kurt Uhlig[a]

[a]Walther-Meissner-Institute, 85748 Garching, Germany

**Abstract.** He-3/He-4 dilution refrigerators (DR) are very common in sub-Kelvin temperature research. We describe a pulse tube precooled DR where a separate He-4 circuit condenses the He-3 of the dilution loop. Whereas in our previous work the dilution circuit and the He-4 circuit were separate, we show how the two circuits can be combined. Originally, the He-4 loop with a base temperature of ~ 1 K was installed to make an additional cooling power of up to 100 mW available to cool cold amplifiers and electrical lines. In the new design, the dilution circuit is run through a heat exchanger in the vessel of the He-4 circuit so that the condensation of the He-3 stream of the DR is done by the He-4 stage. A much reduced condensation time (factor of 2) of the He-3/He-4 gas mixture at the beginning of an experiment is achieved. A compressor is no longer needed with the DR as the condensation pressure remains below atmospheric pressure at all times; thus the risk of losing expensive He-3 gas is small. The performance of the DR has been improved compared to previous work: The base temperature of the mixing chamber at a small He-3 flow rate is now 4.1 mK; at the highest He-3 flow rate of 1.2 mmol/s this temperature increases to 13 mK. Mixing chamber temperatures were measured with a cerium magnesium nitrate (CMN) thermometer which was calibrated with a superconducting fixed point device.

## INTRODUCTION

For experimental applications below a temperature barrier of 0.3 K, DRs are indispensable. In research, cryogen-free DRs which are precooled by pulse tube cryocoolers (PTC) have found widespread use in recent years. PTCs have been commercially available for 13 years; they are especially suited for precooling DRs as there are no moving parts or cold seals in their setup. Therefore, long lifetimes, small vibration amplitudes of the apparatus and little maintenance work are expected. Several commercial cryostat makers offer cryogen-free ("dry") DRs in various sizes; these days, the market share of newly sold cryogen-free DRs is about 80 %.

In the dry DR which we introduced several years ago [1], and in most commercial cryostats there is only one cooling circuit, namely the He-3/He-4 dilution circuit [2]. An additional condensation stage (sometimes referred to as "pot") is not necessary as the condensation of the He-3 flow is taken over by the pulse tube refrigerator. In more recent work, however, we have added a He-4 cooling stage to the DR in order to increase the cooling power of the cryostat at a temperature near 1 K [3]. In our apparatus, DR and 1K-stage are precooled by the same pulse tube refrigerator. It is a commercial cryocooler (cooling power 0.5 W at 4 K); to keep acoustic vibrations low, we chose the smallest model reasonably possible [4].

The refrigeration power of the new 1K-stage reaches up to 100 mW, whereas without it the cooling power of the DR at T ~ 1 K is given by that of the still of the DR which is lower by about a factor of 10. Especially for experimental applications in quantum information technology, increased refrigeration powers are often necessary to cool cold amplifiers and to heat sink electrical cables. The dilution circuit and the 1K-circuit can be installed and operated separately [5]; here we show how these two loops can be combined.

## CRYOSTAT DESIGN

A two-stage PTC is required to precool a DR with 1K-stage. Usually, PTCs are used where the rotary valve and the pulse tubes are mechanically separated from each other and connected only by a flexible pressure hose (Fig.1). Thus the vibrations of the rotary valve and of the compressor are effectively kept away from the cryostat. In all dry

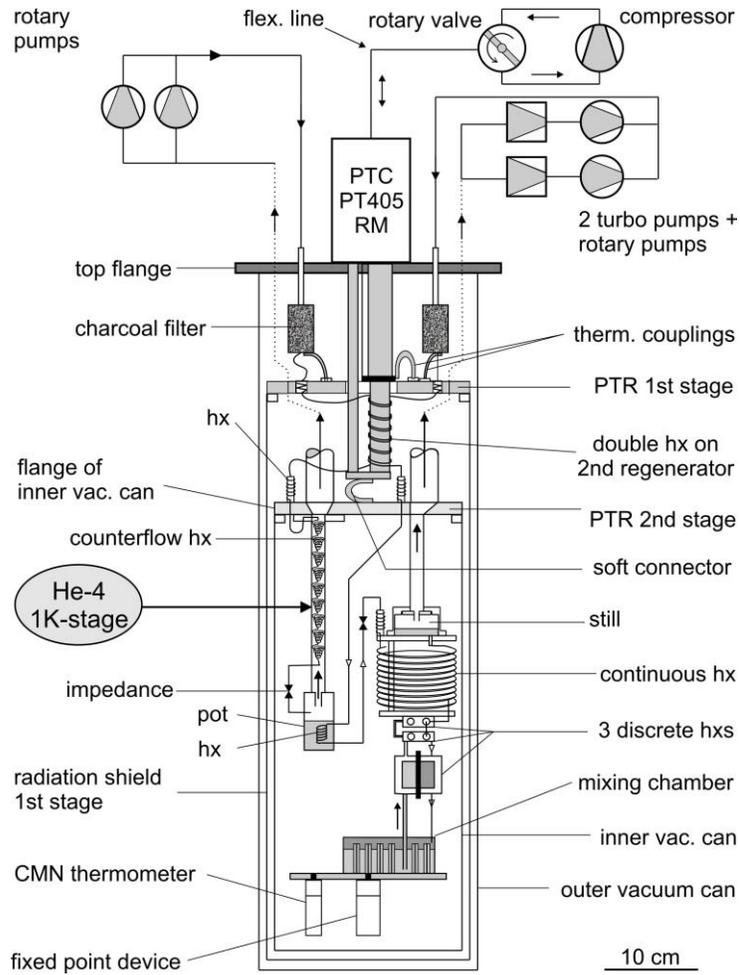

**FIGURE 1.** Cross section of the dry DR with 1K-stage. It is composed of three main components, a PTC, a dilution refrigeration loop and a 1K-circuit. The 1K-stage is on the left side of the figure.

DRs, the first stage of the pulse tube refrigerator is used to cool a radiation shield to a temperature of about 50 K. In our cryostat no superinsulation was wrapped around this shield. Here, the radiational heat load is 17 W which results in a temperature increase of the first stage from 32 K to 50 K. In addition, the first stage cools two charcoal filters to purify the He-3 gas stream of the DR and the He-4 gas flow of the 1K-stage. Charcoal traps with $N_{2,liq}$ are not necessary with this cryostat. The second stage of the pulse tube refrigerator reaches a final temperature near 2.5 K. To measure the temperature of the first stage, a Pt1000 sensor [6] was installed, and for the second stage a Cernox resistance thermometer [7] was available.

Contrary to almost all commercial DRs, ours is equipped with an inner vacuum can. It has an inner diameter of 200 mm and is made from aluminum. It is mechanically connected with the second stage of the PTC and can be filled with exchange gas ($H_2$ or Ne) during cooldown from room temperature to T ~ 15 K. When the dilution refrigeration unit reaches this temperature, the exchange gas can be pumped (e. g. for leak checking) or, alternatively, remain in the inner vacuum can where it freezes upon further cooling. After precooling, the inner vacuum can serves as a radiation shield.

Commercial DR manufacturers don't use the inner vacuum can with their DRs to reach larger inner diameters (~ 40 cm), but have to install either gas switches or an additional He-4 loop to cool the dilution unit from 300 K to 15 K; this gas loop is driven by a separate compressor and has to be carefully evacuated after cooldowns to avoid problems with a superfluid film in the loop which could cause thermal shorts of the DR. Generally, precooling the DR with exchange gas is quicker and more convenient. In the end, the experimental application determines which version is best suitable.

The 1K-stage is constructed like a JT-circuit; the He-4 flow is cooled by the PTC to 2.5 K, further cooled in a counterflow heat exchanger and expanded in a flow restriction. This restriction is made from a thin piece of capillary; therefore, the flow impedance is fixed and not variable like in some commercial He-4 cryostats. After the expansion, the liquid fraction of the flow accumulates in a vessel (V = 53 cm$^3$), whereas the gas fraction is run back to the pumps through the aforementioned counterflow heat exchanger. A simple gas handling system completes the setup of the 1K-stage.

The dilution refrigeration unit has been described before [8]. In contrast to our original construction and to all commercial DRs, there is no counterflow heat exchanger in the pumping line of the still (Fig. 1). Recently, we added a third concrete heat exchanger to the dilution unit to reach a lower base temperature of the mixing chamber. The mixing chamber is big (V = 125 cm$^3$); its bottom plate is made from silver and is thermally connected to a silver sponge so that the thermal boundary resistance between the liquid helium and the silver plate can be bridged. The condensation line of the dilution circuit is run through a heat exchanger which is placed in the vessel of the 1K-stage. Thus the He-3, after being precooled by the PTC, is condensed and cooled to the temperature of the 1K-stage in this heat exchanger. It consists of a capillary which is wound to a spiral of 10 mm i.d.. In earlier experiments this capillary was soft soldered to the outside of the vessel which worked equally well. To measure the temperatures of the various stages of the DR, RuO resistance thermometers [9] were used, and for measuring the lowest temperatures of the mixing chamber a commercial CMN thermometer was available [10].

## PERFORMANCE

The refrigeration power of the 1K-stage is shown in Fig. 2. For thermometry we used a RuO-thermometer and a metal film resistor for heating. To generate the right curve of Fig. 2, a standard rotary pump (Edwards ED80) was used, and for the middle curve of Fig. 2 three rotary pumps were run in parallel (3 x Alcatel 2033H). The lowest temperatures of the 1K-circuit were reached with turbo pumps (two pumps in parallel, 2 x Varian TV-551). The condensation pressure of the circuit was always below 1 bar (Fig. 3) [11].

After a cooldown from room temperature, the condensation rate of the He-3/He-4 gas mixture was 120 std. l/h when the 1K-stage was used. A compressor was not necessary for the condensation process, neither with the 1K-stage nor with the dilution circuit. During condensation, the He-3/He-4 flow produced a heat load of 22 mW onto the 1K-circuit which was transferred by the aforementioned heat exchanger in the vessel of the 1K-stage. He-4 flow and condensation pressure of the 1K-circuit are self-regulating during the condensation. After the condensation of the gas mixture, the heat leak declined as the gas flow decreased. E.g., at a He-3 circulation rate of 1 mmol/s (22.4 std. cm$^3$/s) the heat load into the 1K-stage was 9.4 mW.

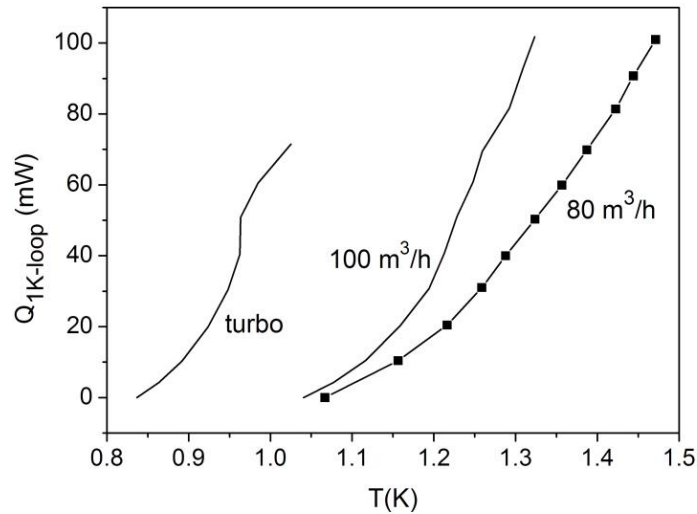

**FIGURE 2.** Refrigeration power of the 1K-stage as a function of its temperature. The two curves on the right of the figure were taken with standard rotary pumps (base temperatures ~ 1.05 K; pumping speeds 80 m$^3$/h and 100 m$^3$/h); the curve on the left was taken with a turbo pump backed by rotary pumps. For details see text.

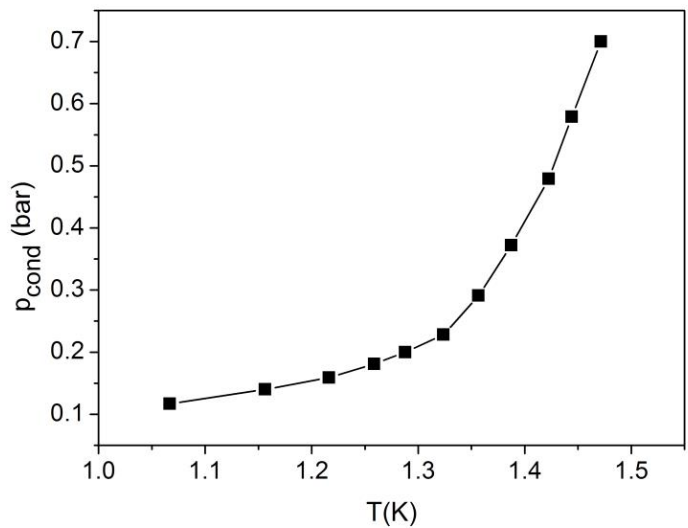

**FIGURE 3.** Condensation pressure of the 1K-loop as a function of its temperature.

If necessary, the He-3/He-4 mash could be condensed without the 1K-circuit in operation [12]. Then the condensation rate was reduced from 120 std. l/h to 55 std. l/h and an inlet pressure of about 5 bar was required; here a small membrane compressor was needed. Now, without the 1K-stage, the circulating He-3 was condensed at the second stage of the PTC (T ~ 2.5 K). At this temperature, the vapor pressure of He-3 is ~ 0.44 bar; for condensation, the inlet pressure has to exceed this value. After the condensation, the He-3 flows towards a flow restriction; the molar volume of He-3 at T ~ 2.5 K is higher as if it was condensed at T ~ 1 K, and thus the flow through the impedance at a given pressure is smaller. Finally, the He-3/He-4 gas could be condensed with the 1K-stage at the beginning of an experiment; then the 1K-stage was shut off and the DR was run without the 1K-stage. This feature was used regularly in the past, especially to run the DR on standby over weekends.

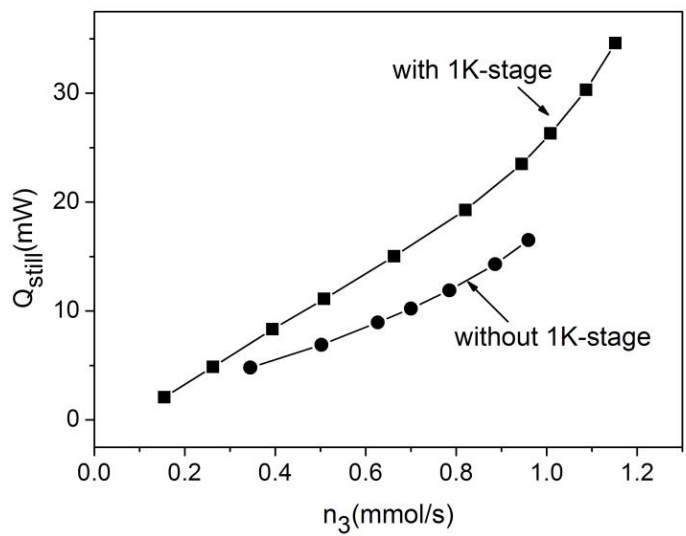

**FIGURE 4.** Cooling power of the still as a function of the He-3 flow rate. For details see text and [11].

If a higher refrigeration power of the 1K-stage is needed, the He-4 circulation rate has to be increased and therefore the pumping speed of the pumps in the 1K-circuit has to be increased, accordingly. Most likely, for higher circulation rates it would be advisable to go with a bigger PTC.

In Fig. 4 the cooling power of the still is depicted as a function of the He-3 circulation rate where the upper curve was taken with the 1K-circuit in operation, and the lower curve when it was not. With the 1K-circuit operating, the enthalpy difference of the incoming He-3 between 2.5 K and 1 K is dumped at the 1K-stage and is therefore available as additional cooling power of the still.

The maximum cooling power of the mixing chamber as a function of the He-3 circulation rate has been given before [11]. The maximum cooling power was about 600 µW at a He-3 flow of 1.1 mmol/s; this corresponds with the maximum flow rate of the turbo pumps used in this particular experiment (2 x Pfeiffer TMH 1601). The pressure at the inlet of the DR was 1 bar. Without the 1K-stage in operation and at an inlet pressure of 1 bar, the He-3 circulation rate of the DR was reduced from 1.1 mmol/s to 0.85 mmol/s. Then the corresponding maximum cooling power was 450 mW.

All the temperatures below 1 K were measured with RuO temperature sensors. In addition, a commercial CMN (cerium magnesium nitrate) thermometer was attached to the bottom plate of the mixing chamber. CMN is a salt whose susceptibility follows a Curie-Weiss law to ultralow temperatures and is therefore used for thermometry purposes. The susceptibility of the CMN sample was measured with an LCR meter (Agilent 4263B) or with a lock-in. For calibration, a superconducting fixed-point-device was available. It consists of five metallic samples with fixed superconducting transition temperatures between 205 mK and 15 mK. The fixed point device we used is an older model from the former NBS; today fixed point devices are made by HDL Inc. [13]. The calibration of the CMN-thermometer with the fixed point device is depicted in Fig. 5.

In Fig. 6 the temperatures of the 1K-stage, of the still and of the mixing chamber are given in dependence of the He-3 throughput. The base temperature of the mixing chamber rose from 4.1 mK at a small He-3 flow to 13 mK at the highest flow rates. Simultaneously the still temperature rose from 0.5 K to 0.8 K, whereas the temperature of the 1K-stage remained almost constant near 1.2 K, independent of the He-3 flow in the DR.

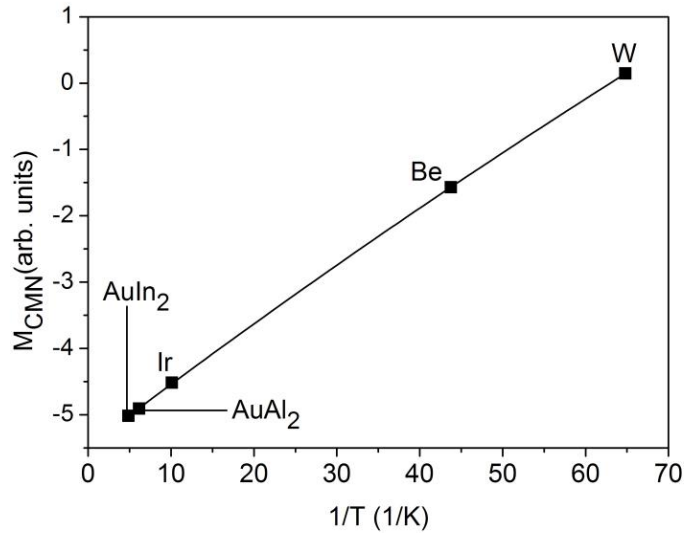

**FIGURE 5.** Magnetization of CMN for the five superconducting transition temperatures of the fixed point device (1/T - plot). A Curie-Weiss-curve is fitted to the data.

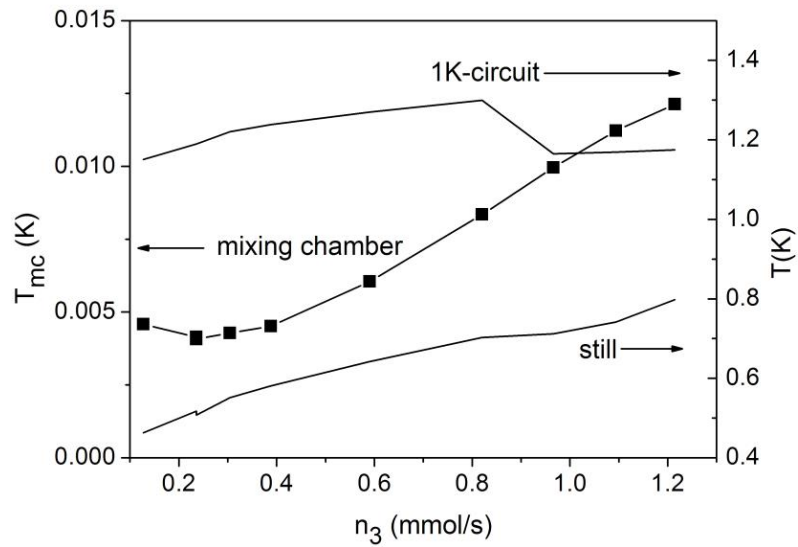

**FIGURE 6.** Base temperature of the mixing chamber as a function of the He-3 flow (center curve, left scale). Also shown are the corresponding temperatures of the still and the 1K-loop (right scale) [11].

## SUMMARY

A cryogen-free DR has been described which differs in several ways from commercial DR cryostats. Instead of a separate precool circuit to cool the dilution unit from room temperature to T ~ 10 K, we use an inner vacuum can which can be filled with exchange gas and also serves as a radiation shield. In the cryostat, a He-4 circuit is installed, in addition to the dilution refrigeration circuit. This He-4 circuit has a base temperature of about 1 K and a cooling power of up to 100 mW. The dilution circuit is thermally coupled to the 1K-circuit via a heat exchanger; with it a speedy condensation of the He-3/He-4 mash is accomplished at the start of an experiment, and in addition the cooling power of the still of the DR is improved.

## ACKNOWLEDGMENTS

The author thanks Oxford Instruments for interest and support.